\documentclass[journal,twoside,web]{IEEEcolor}
\usepackage{generic}
\usepackage{cite}
\usepackage{amsmath,amssymb,amsfonts}
\usepackage{algorithmic}
\usepackage{graphicx}
\usepackage{balance}
\def\BibTeX{{\rm B\kern-.05em{\sc i\kern-.025em b}\kern-.08em
    T\kern-.1667em\lower.7ex\hbox{E}\kern-.125emX}}
\begin{document}
\title{Graphene FET on diamond for high-frequency electronics}
\author{M. Asad, \IEEEmembership{Student Member, IEEE}, S. Majdi, A. Vorobiev, \IEEEmembership{Senior Member, IEEE}, K. Jeppson, \IEEEmembership{Life Senior Member, IEEE}, J. Isberg and J. Stake, \IEEEmembership{Senior Member, IEEE}
\thanks{This project was supported in part by the European Union's Horizon 2020 research and innovation programme (Grant agreement No 881603), in part by the Swedish Research Council (Grant No. 2017-04504) and in part form Swedish Energy Agency, Grant no: 48591-1.}
\thanks{M. Asad, A. Vorobiev, K. Jeppson and J. Stake are with Chalmers University of Technology, Department of Microtechnology and Nanoscience, SE-412 96 Gothenburg, Sweden. S. Majdi and J. Isberg are with Division for Electricity, Department of Electrical Engineering, Uppsala University, Box 65, SE-751 03 Uppsala, Sweden.
(e-mail: saasad46@gmail.com).}}
\maketitle

\begin{abstract}
Transistors operating at high frequencies are the basic building blocks of millimeter-wave communication and sensor systems. The high charge-carrier mobility and saturation velocity in graphene can open way for ultra-fast field-effect transistors with a performance even better than what can be achieved with III-V-based semiconductors. However, the progress of high-speed graphene transistors has been hampered by fabrication issues, influence of adjacent materials, and self-heating effects. Here, we report on the improved performance of graphene field-effect transistors (GFETs) obtained by using a diamond substrate. An extrinsic maximum frequency of oscillation $f_\mathrm{max}$ of up to 54 GHz was obtained for a gate length of 500 nm. Furthermore, the high thermal conductivity of diamond provides an efficient heat-sink, and the relatively high optical phonon energy of diamond contributes to an increased charge-carrier saturation velocity  in the graphene channel. Moreover, we show that GFETs on diamond exhibit excellent scaling behavior for different gate lengths. These results promise that the GFET-on-diamond technology has the potential of reaching sub-terahertz frequency performance.
\end{abstract}


\section{Introduction}
\label{sec:introduction}
Graphene is a 2D material with unique electrical properties such as extremely high charge-carrier velocity useful for transit types of devices \cite{novoselov2004}. There has been extensive research and progress on high-frequency graphene electronics since the first top-gated graphene field-effect transistor (GFET) was demonstrated in 2007 \cite{lemme2007}. In 2012, excellent GFET performance with intrinsic transit frequencies above 400 GHz was achieved for a 67-nm gate length GFET \cite{cheng2012}. However, this excellent performance was not matched by the maximum frequency of oscillation that lagged considerably due to non-negligible gate resistances.
\begin{figure}
\centerline{\includegraphics[width=0.9\columnwidth]{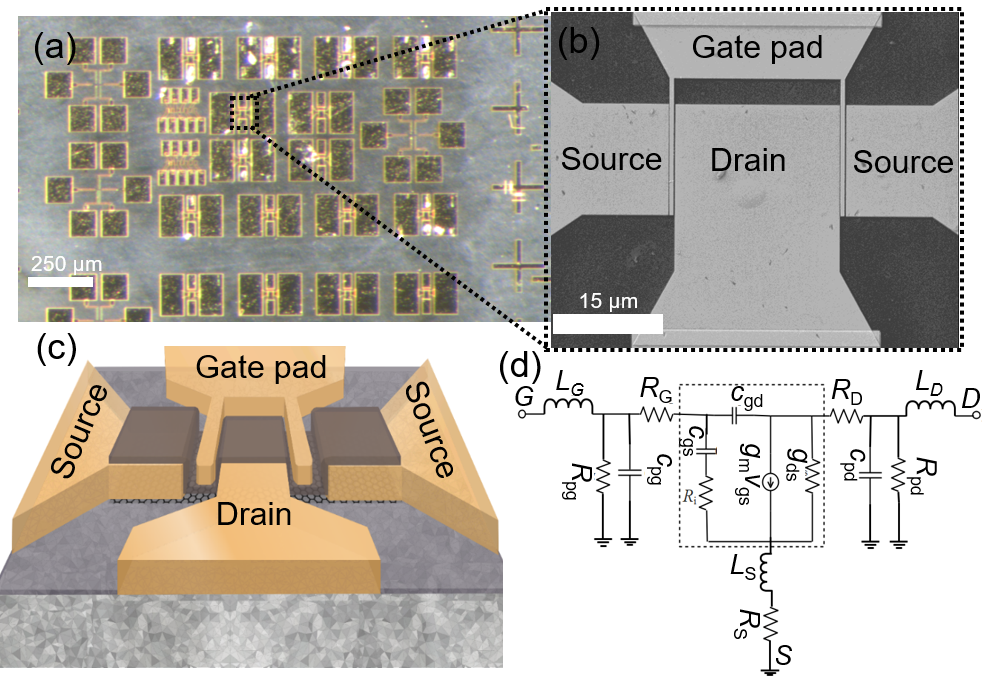}}
\caption{ Views of top-gate, dual-channel, high-frequency GFET on single crystal diamond substrate: (a) Photograph of the test chip, (b) SEM image, (c) schematic cross-sectional view, (d) small-signal equivalent circuit and  dashed region indicates the intrinsic transistor elements.}
\label{fig1}
\end{figure} 

In GFETs, the maximum frequency of oscillation is in part limited by poor drain current saturation which results in a non-negligible drain output conductance \cite{schwierz2013}. In 2016, an extrinsic transit frequency ($f_\mathrm{T}$) of 50 GHz and a maximum oscillation frequency ($f_\mathrm{max}$) of 40 GHz was achieved for a gate length of 200 nm using quasi-freestanding bilayer epitaxial graphene grown on a SiC (0001) substrate \cite{yu2016}. Recently, using an improved fabrication process for CVD GFETs, Bonmann \emph{et al.} demonstrated an extrinsic $f_\mathrm{T}$ of 34 GHz and a matching $f_\mathrm{max}$ of 37 GHz for 500-nm GFETs with promising scaling behaviour \cite{bonmann2018}. In order to minimise the output conductance it is essential to minimize the number of charge carriers not induced by the field such as carriers due to contaminants and traps in the adjacent materials, or carriers induced by self-heating.
 
Another performance-limiting factor is the influence on the charge-carrier velocity due to optical-phonon scattering with the materials surrounding the graphene channel. Hence, surrounding materials with high optical-phonon energies are preferred since there is a direct correlation between the charge-carrier velocity and the transit frequency as confirmed by ref \cite{9316726}.
  
In this work, we utilize both the high surface optical phonon energy and the high thermal conductivity of diamond as a substrate to increase the GFET performance. A record high extrinsic $f_\mathrm{max}$ of 54 GHz was achieved for a top-gated GFET on a single-crystal diamond substrate. We estimate the charge-carrier saturation velocity being as high as $3.2 \cdot 10^7$ cm/s. Finally, we show that the high-frequency performance scales with the gate length, which indicates that GFETs on diamond have the potential of reaching sub-terahertz frequencies.
\begin{figure}
\centerline{\includegraphics[width=1\columnwidth]{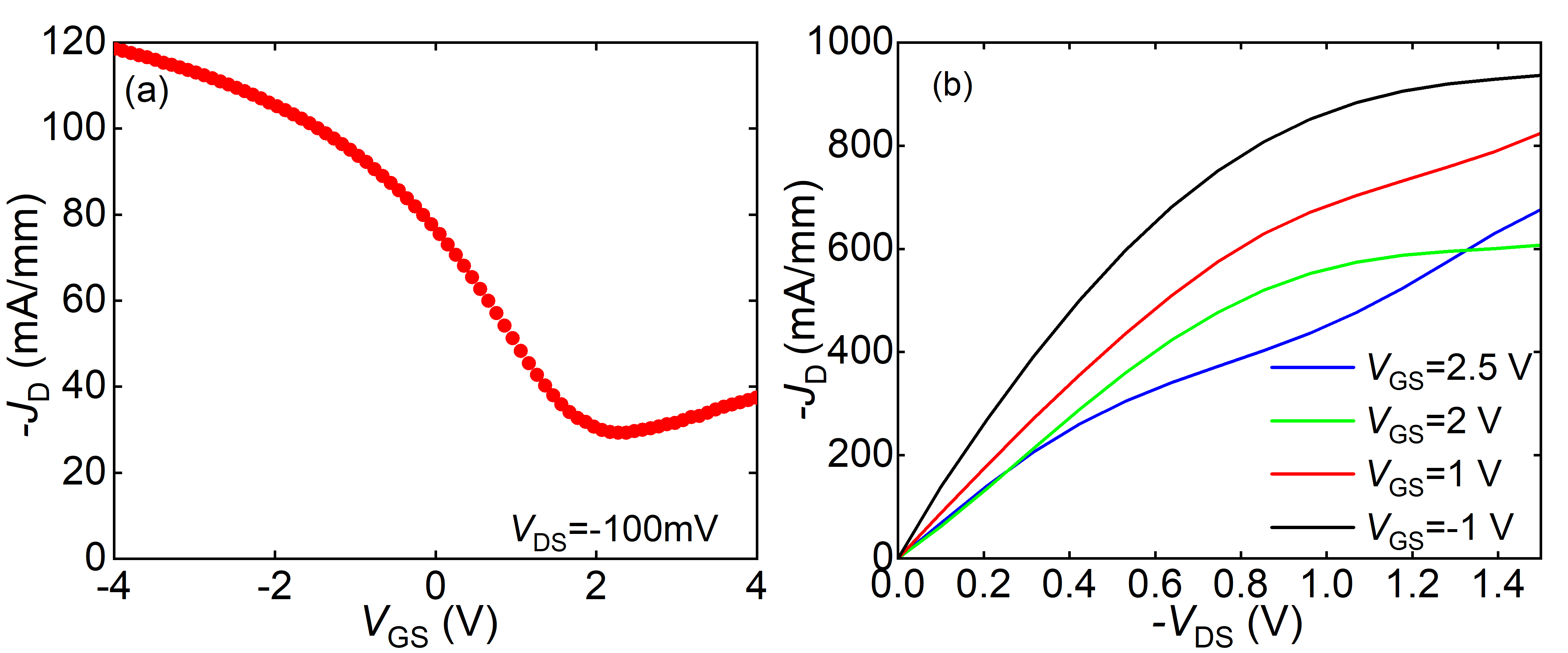}}
\caption{GFET transfer characteristics (a) and output characteristics. (b) GFET geometry $L_\mathrm{g}$ =0.5 $\mu$m, W=30 $\mu$m.} 
\label{IV}
\end{figure}
\begin{figure}
\centerline{\includegraphics[width=0.7\columnwidth]{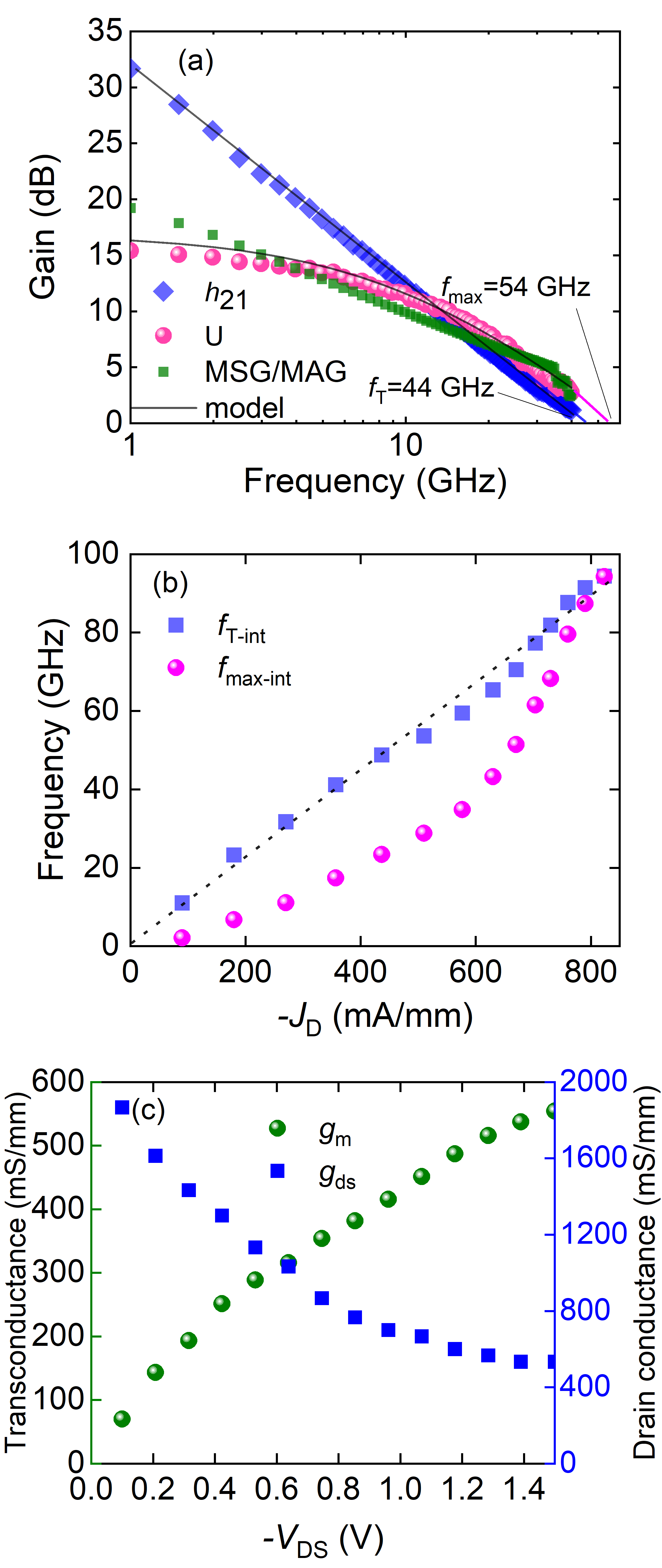}}
\caption{Graphs showing (a) small-signal current gain $h_\mathrm{21}$, maximum stable gain/maximum available gain (MSG/MAG), and Mason’s unilateral gain $U$ versus frequency obtained at $V_\mathrm{GS}$=1 V and $V_\mathrm{DS}$=-1.5 V. Extrapolated extrinsic transit and maximum oscillation frequencies of 44 and 54 GHz, respectively, are obtained. Solid lines are obtained from the S-parameters simulated using small-signal equivalent circuit analysis. (b) Intrinsic transit and maximum-oscillation frequencies versus drain current density, (c) normalized small-signal conductances $g_\mathrm{m}$ and $g_\mathrm{ds}$ versus drain voltage.}
\label{fig2}
\end{figure}
\section{Methods}
Top-gated dual-channel RF GFETs were fabricated on a high-quality free-standing single-crystal diamond substrate. The commercially available diamond substrate was homoepitaxially grown in the (100) direction using chemical vapor deposition (CVD) by Element Six Ltd \cite{isberg2002}. Fig. \ref{fig1}(a) shows an optical micrograph of a set of GFETs with gate lengths ($L_\mathrm{g}$) varying from 0.5 $\mu$m to 2 $\mu$m and a total gate width of 30 $\mu$m. Fig. 1(b) shows a SEM image of a 0.5 $\mu$m two-finger GFET. For illustration, Fig. \ref{fig1}(c) shows a three-dimensional view of the GFET layout. As a first step of the fabrication process, monolayer CVD graphene was transferred on to the diamond substrate, followed by deposition of a 5-nm thick $\mathrm{TiO_2}$/Ti protection layer for avoiding contact between the electron-beam resist defining the source/drain areas and the graphene. After wet-etch removal of the protection layer in the source/drain resist openings, Ti/Pd/Au (1 nm/15 nm/285 nm) was deposited and source/drain contacts were fabricated using lift-off. The use of a protection layer is important for obtaining a clean metal/graphene interface and a low contact resistance. Next, the graphene mesa was defined. After removal of the remaining protection layer from the channel area, the gate dielectric stack was formed by a 5-nm thick thermally oxidized $\mathrm{Al_2O_3}$ seed layer and an 18-nm thick $\mathrm{Al_2O_3}$ top layer deposited by atomic layer deposition. The gate fingers, the gate pad, and the source/drain pads were formed by deposition of a Ti/Au (100 nm/300 nm) layer. All patterns were defined using electron-beam lithography.
 
For GFET characterization, transfer and output characteristics were obtained using a Keithley 2612B dual-channel source meter. High-frequency $S$-parameters were measured up to 40 GHz using an Agilent N5230A network analyzer. Two-port open-short-load-through calibration was performed using a standard calibration chip prior to the $S$-parameter measurements. De-embedding and extraction of the small-signal equivalent-circuit elements was performed based on the method described by Dambrine et al. \cite{dambrine1988}. Special test structures were included to find the pad capacitances and inductances from the S-parameter measurements. For analysing the GFET high-frequency performance, the small-signal equivalent circuit shown in Fig. \ref{fig1}(d) was used.
\section{Results and discussions}
Fig. \ref{IV} shows the GFET transfer and output characteristics indicating typical GFET behavior. The $J_\mathrm{D}$ vs. $V_\mathrm{DS}$ curve obtained for $V_\mathrm{GS}$=1 V shows the conditions under which the high-frequency performance was measured. Fig. \ref{fig2} summarizes the measurement results obtained for the same GFET biased for maximum gain.
First, the frequency-dependent current gain ($h_\mathrm{21}$), the Mason's unilateral power gain ($U$), and the maximum stable gain/maximum available gain (MSG/MAG) obtained from the $S$-parameters are shown in Fig. \ref{fig2}(a). An extrinsic $f_\mathrm{T}$ as high as 44 GHz and an extrinsic $f_\mathrm{max}$ as high as 54 GHz are indicated by the figure. To the best of our knowledge, this is the highest reported extrinsic performance of top-gated CVD GFETs so far \cite{guo2013,lin2010,wu2012,lyu2016,bonmann2018,9316726}. Fig. \ref{fig2}(b) shows an almost linear relationship between the intrinsic transit frequency $f_\mathrm{T-int}$ and the drain current density $J_\mathrm{D}$, indicating a practically constant charge-carrier density $n$ throughout the whole drain bias range (0 to 1.5 V). This conclusion is based on the simple first-order assumption of an intrinsic transit frequency $f_\mathrm{T-int}$=$v_\mathrm{d}/(2\pi L_\mathrm{g})$ and a drain current density $J_\mathrm{D}$=$qnv_\mathrm{d}$, where $v_\mathrm{d}$ is the drift velocity, and $q$ is the elementary charge. Based on these assumptions an effective charge carrier concentration $n$ = 1.8$\cdot10^{12}$ $\mathrm{cm^{-2}}$ was obtained. Furthermore, the assumption of a constant charge-carrier density leads to the conclusion that the charge-carrier drift velocity profile can be obtained not only from the $f_\mathrm{T-int}$ vs. $V_\mathrm{DS}$ (as in previous work \cite{9316726}) but also directly from the output characteristic $J_\mathrm{D}$ vs. $V_\mathrm{DS}$. Finally, we can conclude that the effect of self-heating, typically resulting in an increasing charge-carrier density, is sufficiently low due to the high thermal conductivity of the diamond substrate.
\begin{figure}
\centerline{\includegraphics[width=0.7\columnwidth]{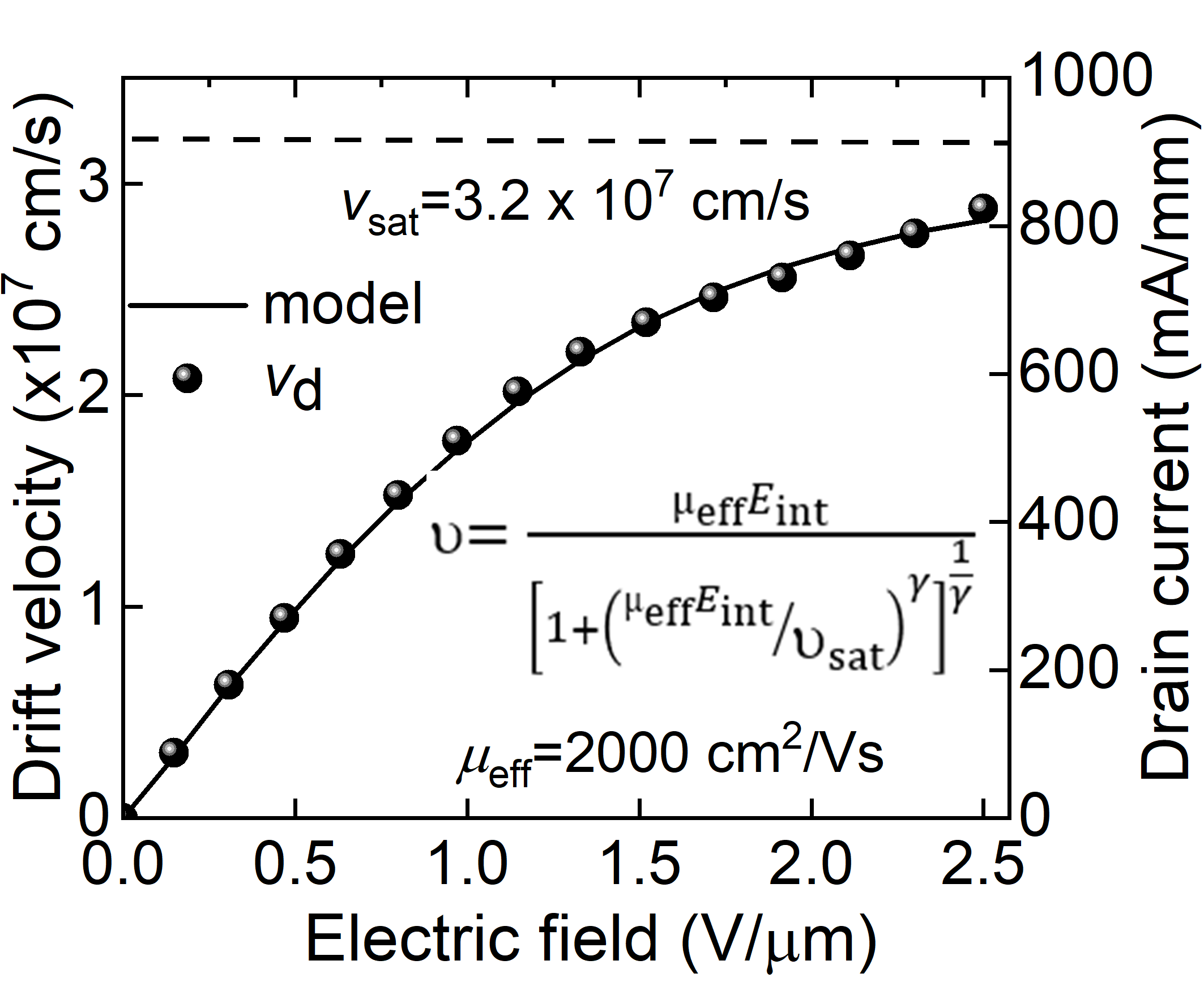}}
\caption{ Charge-carrier velocity (left axis) versus intrinsic electric field along the channel derived from the drain current density $J_\mathrm{D}$ vs. $E_\mathrm{int}$ (right axis) by assuming a constant charge-carrier concentration n=1.8$\cdot10^{12}$ $\mathrm{cm^{-2}}$ obtained from the slope of the $f_\mathrm{T-int}$ vs $J_\mathrm{D}$ graph shown in Fig. \ref{fig2}b. The Caughey-Thomas model \cite{caughey1967} (solid line) with $\gamma$=3 was used to estimate the saturation velocity (dashed line).}
\label{fig3}
\end{figure}
From the broadband $S$-parameter measurements, the bias-dependent small-signal equivalent circuit parameters $g_\mathrm{m}$, $g_\mathrm{ds}$, $C_\mathrm{gs}$, and $C_\mathrm{gd}$ were extracted and normalized by the gate width. The values of $C_\mathrm{gs}$ and $C_\mathrm{gd}$ are 500$\pm$50 fF/mm, 450$\pm$50 fF/mm, respectively, and are almost constant across the drain bias range. Figs. \ref{fig2}(c) shows conductances $g_\mathrm{m}$ and $g_\mathrm{ds}$ versus the extrinsic drain voltage $V_\mathrm{DS}$. The observed increase in transconductance $g_\mathrm{m}$ and decrease in drain conductance $g_\mathrm{ds}$ with $V_\mathrm{DS}$ are the two main factors for the enhanced high-frequency performance of these devices. 
Other extrinsic circuit elements such as the pad inductance ($L_\mathrm{G}$= 13.8 pH, $L_\mathrm{S}$=1.6 pH, $L_\mathrm{D}$=5 pH), source/drain resistance ($R_\mathrm{S}$/$R_\mathrm{D}$=23 $\Omega$), gate resistance, ($R_\mathrm{G}$=8 $\Omega$), parasitic pad capacitances ($C_\mathrm{PG}$ =15 fF, $C_\mathrm{PD}$= 4 fF), and parasitic pad resistances ($R_\mathrm{pad}$=21 k$\Omega$) are assumed to be bias-independent. Typically, due to zero-bandgap in monolayer graphene, GFETs reveal a linear output characteristic without drain current saturation resulting in a relatively high $g_\mathrm{ds}$ which has an adverse impact on the power gain i.e., limiting the $f_\mathrm{max}$. In this work, high current density due to high charge carrier velocity and significant drain current saturation due to velocity saturation are observed and verified by the drain current modeling. The drain current saturation results in low $g_\mathrm{ds}$ and, thus enhanced $f_\mathrm{max}$. Most promising way of increasing the high-frequency performance of GFET is by increasing the $g_\mathrm{m}$, which is proportional to the $v_\mathrm{d}$. The transconductance $g_\mathrm{m}$ shown in Fig. \ref{fig2}(d) increases with the $V_\mathrm{DS}$ and shows signs of saturation at higher field. 

\begin{figure}
\centering
\includegraphics[width=0.7\columnwidth]{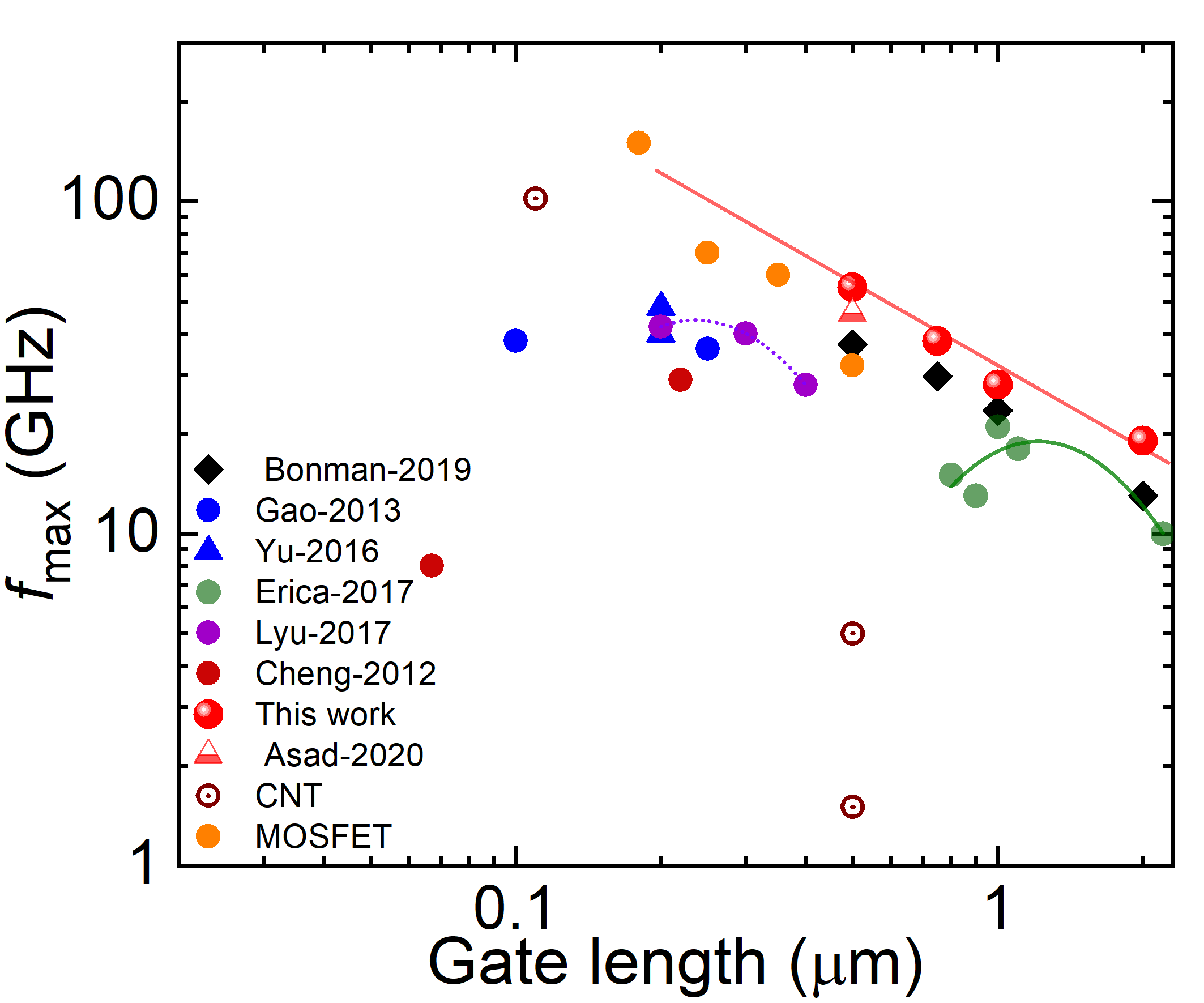}
\caption{Maximum frequency of oscillation ($f_\mathrm{max}$) versus gate length ($L_\mathrm{g}$). Solid red circles
represent data from this work and the solid line showing a 1/$L_\mathrm{g}$ dependence demonstrates the
device scaling behavior. Also shown, for comparison, is data from GFETs on CVD, exfoliated and epitaxial graphene on silicon and SiC substrates \cite{cheng2012,guo2013,lyu2016,guerriero2017,bonmann2018,9316726}, from carbon nanotube FETs \cite{wang2014,rutherglen2019} and from MOSFETs of similar gate lengths \cite{de1995,tiemeijer2001,Johnson1998}.}
\label{fig5}
\end{figure}
Fig. \ref{fig3} shows the charge-carrier velocity found from the dependence of the drain current on the intrinsic drain field. The saturation velocity $v_\mathrm{sat}$ was derived by using the Caughey-Thomas velocity model \cite{caughey1967} assuming a constant effective charge-carrier density – as previously concluded from the almost linear relationship between $f_\mathrm{T}$ versus $J_\mathrm{D}$ across the drain voltage range shown in Fig. \ref{fig2}(b). The $v_\mathrm{sat}$ in the diamond GFET is estimated being as high as 3.2 $\cdot$ 10$^7$ cm/s. For comparison, the $v_\mathrm{sat}$ in an hBN-encapsulated graphene Hall-bar test structure was reported to 5$\cdot$ 10$^7$ cm/s \cite{yamoah2017}.
 
To demonstrate the future prospects of the diamond GFET technology, the extrinsic performance of GFETs for different gate lengths is compared with published results in Fig. \ref{fig5}. In this work, the performance follows the solid trendline $1/L_\mathrm{g}^{0.9}$, typical for Si-MOSFET and HEMT technologies \cite{schwierz2013}, which indicates a good scaling behaviour. For instance, we foresee that $f_\mathrm{max}$ $>$ 100 GHz could be achieved for 200-nm GFET devices. The main improvements in the high-frequency performance and in the scaling behavior of the extrinsic $f_\mathrm{max}$ are mainly attributed to a higher carrier velocity, improved current saturation, and reduced effects of self-heating.

\section{Conclusions}
In summary, the potential of top-gated graphene field-effect transistors on diamond substrate has been demonstrated. Substantial improvements in scaling behaviour and in maximum frequency of oscillation have been achieved due to the unique properties of diamond such as high surface-optical-phonon energy and high thermal conductivity when being used as a substrate. To reach the full potential of GFETs on diamond, both  material quality and fabrication processes must be improved to achieve higher charge-carrier mobilities and saturation velocities. 
\clearpage
\balance
\bibliographystyle{IEEEtran}
\bibliography{IEEEfull,asd}
\end{document}